\documentclass[12pt]{article}
\usepackage{epsf}
\usepackage{relsize}
\def\babar{\mbox{\slshape B\kern-0.1em{\smaller A}\kern-0.1em
    B\kern-0.1em{\smaller A\kern-0.2em R}}}
\begin{document}

\begin{titlepage}

\rightline{ISU-HET-02-1}
\rightline{hep-ph/0201159}
\rightline{June, 2002}

\begin{center}

{\Large\bf Probing Intrinsic Charm with Semileptonic \mbox{\boldmath
$B$} Decays}

\medskip

\normalsize
{\large F. Gabbiani$^{}$, Jianwei Qiu$^{}$, and G. Valencia$^{}$} \\
\vskip .3cm
$^{}$Department of Physics and Astronomy, Iowa State University,
Ames, Iowa 50011\\
\vskip .3cm

\end{center}

\begin{abstract}

We discuss semileptonic $B$ decays of the form $B \rightarrow$
$J/\Psi$ $e \nu$ $X$ as possible probes of intrinsic charm. We
calculate the leading order perturbative contribution to the process
$B$ $\rightarrow$ $J/\Psi$ $e^- \overline{\nu}_e$ $X$ and find it to be
unobservably small, with a branching ratio $\sim$
10$^{-10}$. We propose a modified spectator model to estimate the intrinsic 
charm contribution and find that it can be significantly larger, with 
a branching ratio for $B$~$\rightarrow$~$(c\bar{c})$~$e^-
\overline{\nu}_e$~$X$ as large as $5\times 10^{-7}$. We show that the
process could be observed at these levels by CDF assuming a Run II
integrated luminosity of 15~fb$^{-1}$, making this a useful reaction to 
probe the idea of intrinsic charm.

\end{abstract}

\end{titlepage}

\section{Introduction}

Interesting possibilities open up if we consider wavefunctions of
hadronic bound states that contain light-cone Fock states of
arbitrarily high particle number. The higher light-cone Fock
components arise as quantum fluctuations suppressed by $M^2$, where
$M$ is the mass of the fluctuation. The $c \overline{c}$ component of
hadronic Fock states is referred to as intrinsic charm (IC) \cite{IC}.

The possibility of detecting signatures of IC in
$B$-meson physics has been recently discussed in Ref. \cite{hou}.
The authors propose IC as the explanation for a ``slow'' (low
momentum) bump in the inclusive $B$ $\rightarrow$ $J/\Psi \, X$ spectrum
and the softness of the $J/\Psi$ spectrum in $\Upsilon$ $\rightarrow$
$J/\Psi \, X$ decays, contrary to expectations from the color octet
mechanism.
IC appearing in higher light-cone Fock components of the $B$-meson
wavefunction can manifest itself in two possible ways. It can
operate virtually, in a mediation r\^ole, and affect decay processes
by providing additional channels for the weak interactions. In this
way one may enhance some CKM suppressed $B$ decays. This has
been discussed in Ref.~\cite{alex} using a phenomenological mixing
angles approach, and more recently in Ref.~\cite{bg}. Another instance 
of this mediation is given by the hypothesis that the IC component 
of $\rho$ mesons is able
to explain the ``$\rho \pi$ puzzle'' \cite{bk}. IC can also
manifest itself in processes in which the (intrinsic) $c \overline{c}$ 
produces charmed hadrons in the final state. An example of this is 
discussed in Ref. \cite{hsv} to account for charm production
in deep inelastic scattering. Here we consider another example of
the latter 
by looking for $J/\Psi$ production from a $c \overline{c}$ component 
in the $B$ meson wavefunction. 

\begin{figure}[!htb]
\begin{center}
\epsfxsize=16.5cm
\centerline{\epsffile{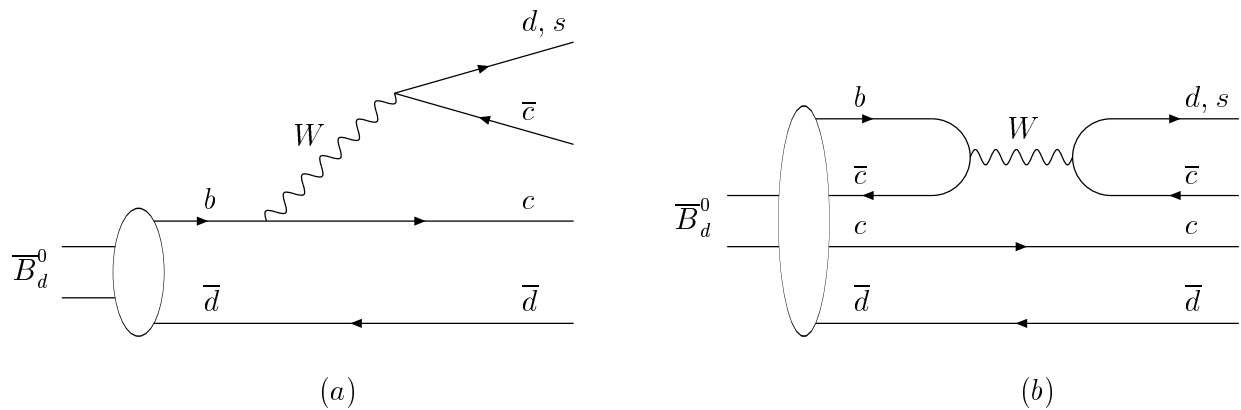}}
\end{center}
\caption{Perturbative contribution to
the decay $\overline{B}^0_d \rightarrow J/\Psi \, X$. (a) is a standard
tree-level graph, (b) is an annihilation graph involving IC in the
initial state. An IC exchange graph is also allowed.}
\label{fig:graph0}
\end{figure}

To determine the process that is best suited for our purpose, 
we first consider purely hadronic decays of the $B$ meson
with just one $c \overline{c}$ pair
in the final state. It is then easy to see from 
Fig. \ref{fig:graph0}, that hadronic channels with a $c\overline c\;
X_{u,d,s}$
quark content are also produced at tree level via $V^{\phantom{l}}_{cb}$, 
with no need for IC. We conclude that these processes are ill-suited 
to study the effects of IC, which would be swamped by a formidable 
background. 
Next, we consider a process with three charm quarks (or antiquarks)
in the final state such as $B
\rightarrow J/\Psi D^{(*)}$ or $B \rightarrow J/\Psi D^{(*)} X$.
The former has been discussed in Ref. \cite{hou} and 
in the latter the phase space available closes rapidly,
suppressing the rates to unobservable levels.

This leads us to concentrate in semileptonic modes with a $J/\Psi$ 
in the final state. These modes have the additional advantage of
presenting
an extremely clean experimental signature containing three charged leptons.
The simplest mode of this type is $B \rightarrow J/ \Psi\; e^-
\overline{\nu}_e \, X$.\footnote{For definiteness we concentrate here on 
channels with a $J/\Psi$ in the final state, but our analysis can 
also be applied to channels with other 
$c\overline{c}$ resonances or with $D \overline{D}$ pairs instead.}
Detectors such as CDF operating at hadron machines,
or experiments at $B$ factories such as \babar\ at SLAC should be able
to identify the three outgoing charged leptons easily. In fact, CDF 
will be looking for the mode $B_c \rightarrow J/ \Psi\; e^-
\overline{\nu}_e$ which has the same topology (and which would
constitute a background to our process). It would therefore be natural 
for them to search for $B \rightarrow J/ \Psi\; e^-\overline{\nu}_e\; X$ 
simultaneously.

This paper is organized as follows: in Section \ref{pert} we calculate
the perturbative QCD contributions to semileptonic processes such as
$B^- \rightarrow J/\Psi \; e \nu \, X$ and show that they are
unobservably small. In Section \ref{intic} we propose a modified
spectator model for the computation of the IC contributions to these
same semileptonic processes, and find that they can be substantially
larger, with branching ratios as large as $5 \times 10^{-7}$. Finally
in Section \ref{conc} we estimate the reach of near future experiments
for these reactions and show that they could observe the mode $B^-
\rightarrow J/\Psi \; e \nu \, X$ if the IC component of $B$ mesons is
as large as has been speculated recently \cite{bg}.

\section{The perturbative calculation}
\label{pert}

\begin{figure}[!htb]
\begin{center}
\epsfxsize=15cm
\centerline{\epsffile{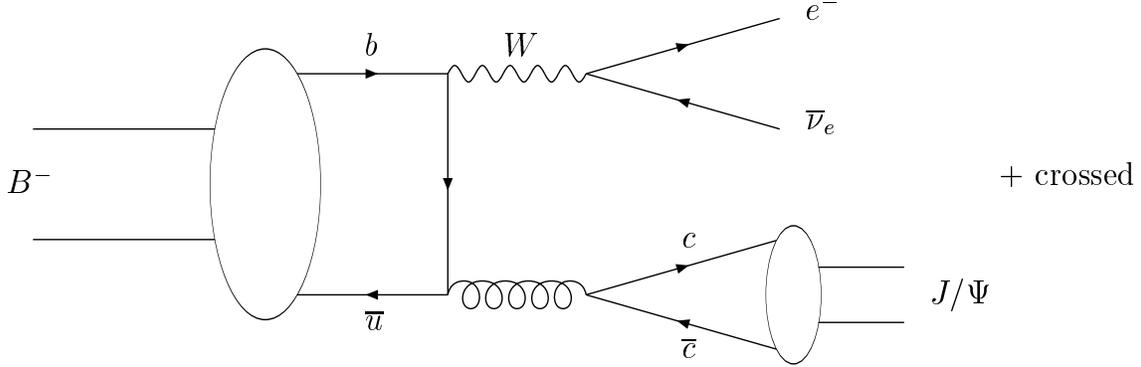}}
\end{center}
\caption{Graphs describing the perturbative contribution to
the decay $B^- \rightarrow J/\Psi \; e^- \overline{\nu} \, X$.}
\label{fig:graph1}
\end{figure}

In this section we compute the perturbative contribution to the decay rate 
of the process $B^- \rightarrow J/\Psi\; e^- \overline{\nu}_e \, X$ from
the two lowest order diagrams in Fig.~\ref{fig:graph1}.  We present 
calculations of the decay rate in both Non-Relativistic QCD (NRQCD)
\cite{BBL-NRQCD} and in the Color Evaporation Model (CEM)\cite{CEM}. We
also derive the leading order contribution to the total inclusive 
$B$-to-($c \bar c)$ semileptonic decay rate.

A hadronic $B$ decay depends on the momentum scale of the meson
wavefunction (1/fm~$\sim\Lambda_{\rm QCD}$) and is non-perturbative in
principle.
Nevertheless, the mass of the heavy quark provides a large momentum scale 
and to the extent that 
$m^{\phantom{l}}_{J/\psi}\sim 2m_c \gg \Lambda_{\rm QCD}$ and 
$m^{\phantom{l}}_b-m^{\phantom{l}}_{J/\psi} \gg \Lambda_{\rm QCD}$, 
the $c\bar{c}$ pair production occurs at short distance and is perturbative. 
On the other hand, the formation of $J/\Psi$ mesons from the initially
compact $c\bar{c}$ pairs takes a long time and is non-perturbative.  
The quantum interference between the production of the $c\bar{c}$ pair
and the formation of $J/\Psi$ mesons should be suppressed by powers of
the ratios of the typical momentum scales of these two processes, such
as $\Lambda_{\rm QCD}^2/m_{J/\psi}^2$ and 
$\Lambda_{\rm
QCD}^2/(m^{\phantom{l}}_b-m_{J/\psi})^2$. Therefore, if
we neglect these power corrections, the decay rate can be factorized as 
\begin{eqnarray} 
\Gamma^{\phantom{l}}_{B^-\rightarrow J/\psi \; e\bar{\nu} \,X}
&=& \sum_{n}
\int dq^2 \int d^4Q \left(
  \frac{d\Gamma^{\phantom{l}}_{B\rightarrow [c\bar{c}]_n \; e\bar{\nu}
\, X}}{d^4Q}\right)
\nonumber \\
&\times &
\delta\left(q^2+(k_c-k_{\bar{c}})^2\right)\,
F_{[c\bar{c}]_n \rightarrow J/\psi}(q^2) ,
\label{fact}
\end{eqnarray}
where $\sum_{n}$ runs over all possible singlet and octet
color states, and $Q^\mu=k_c^\mu + k_{\bar{c}}^\mu$ is the
total momentum of the produced $c\bar{c}$ pair.  In Eq.~(\ref{fact}),
$q^2=(2\vec{k}_c)^2$ is the square of the relative momentum between the
$c$ and the $\bar{c}$ in their rest frame. 
Treating the $c$ or $\bar{c}$ approximately as being 
on their mass-shell,
$k_c^2 = k_{\bar{c}}^2=m_c^2$, $q^2=(k_c-k_{\bar{c}})^2 = Q^2 -
4\, m_c^2$.  The non-perturbative quantity 
$F_{[c\bar{c}]_n \rightarrow J/\psi}(q^2)$ denotes 
the transition probability for a $[c\bar{c}]$ state with 
relative momentum squared $q^2$ to evolve into a physical $J/\psi$
meson. Different choices of
$F_{[c\bar{c}]_n \rightarrow J/\psi}(q^2)$ lead to different
models of $J/\Psi$ formation.

\subsection{Nonrelativistic QCD}
\label{secnrqcd}

The NRQCD formalism for quarkonium production exploits the fact that 
the heavy quark velocity $v_c$ is much less than one, or equivalently 
that the heavy quark three-momentum is much smaller than its mass 
($|\vec{k}_c| \ll m_c$) \cite{BBL-NRQCD}. 
By expanding the $q^2$-dependence of 
$d\Gamma_{B^-\rightarrow [c\bar{c}]_n X}/d^4Q$ around $q^2=0$ 
in Eq.~(\ref{fact}), we effectively derive the leading term in the NRQCD
formalism:
\begin{equation}
\Gamma(B^- \rightarrow J/\Psi\; e^- \overline{\nu}_e \, X)
= \sum_{n}
  \Gamma(B^- \rightarrow [c\bar{c}]_n\, e^- \overline{\nu}_e \, X)\,
  \langle {\cal O}_n^{J/\Psi}\rangle\, .
\label{fac-NRQCD}
\end{equation}
The constants $\langle {\cal O}_n^{J/\Psi}\rangle
\propto \int dq^2 F_{[c\bar{c}]_n \rightarrow J/\psi}(q^2)$ are
matrix elements of local operators in coordinate space,   
often called NRQCD matrix elements. They contain  
all the non-perturbative information on
the $J/\Psi$ formation from the $c\bar{c}$ pair. 
According to the NRQCD formalism for heavy quarkonium production,
the color of the gluon in Fig.~\ref{fig:graph1} is neutralized by
non-perturbative soft gluons in the production process, which are not
explicitly shown.
For the lowest order diagrams in Fig.~\ref{fig:graph1} only a spin-1
and color-octet $c\bar{c}$ state is able to produce a $J/\Psi$,
and the corresponding matrix element is
$\langle {\cal O}_8^{J/\Psi}(^3S_1)\rangle$ \cite{mal}.

To the extent that the $J/\Psi$ mass is large, the partonic process 
in the decay amplitudes 
of $B^- \rightarrow [c\bar{c}]_8\, e^- \overline{\nu}_e \, X$ in
Fig.~\ref{fig:graph1} takes place at short distance
$\sim 1/m^{\phantom{l}}_{\Psi}$. More precisely, the gluon is off-shell by
the $J/\Psi$ mass, and the denominator of the $t$- or $u$-channel fermion
propagator is at least 5~GeV$^2$. Therefore, the decay in 
Fig.~\ref{fig:graph1} can be factorized into a 
hadronic matrix element for $B\rightarrow b\bar{u}$, and 
a partonic process $b \bar{u} \rightarrow [c\bar{c}]_8\, e^- \overline{\nu}_e$.
In order to achieve this factorization we formally
write the decay amplitudes of $B\rightarrow g^{\star} W^{\star}$
in Fig.~\ref{fig:graph1} as:
\begin{eqnarray}
M^{\alpha\mu}_i = \int {{d^4 p^{\phantom{l}}_b} \over {(2 \pi)^4}}\,
{\rm Tr}\left[ \widehat{T}^{\phantom{l}}_B(p^{\phantom{l}}_B, \;
p^{\phantom{l}}_b)
\cdot \widehat{H}^{\alpha\mu}_i(p^{\phantom{l}}_b, \;
p^{\phantom{l}}_{\Psi}, \; q)
\right] , \qquad i=1,\;2 \;,
\label{M-TH}
\end{eqnarray}
where $\widehat{T}^{\phantom{l}}_B$ and $\widehat{H}$ represent 
the matrix elements of $B\rightarrow b\bar{u}$ and the partonic process 
$b\bar{u}\rightarrow g^{\star} W^{\star}$ respectively. 
The partonic momenta satisfy  
$p^{\phantom{l}}_{\bar{u}}=p^{\phantom{l}}_{B}-p^{\phantom{l}}_b$,
and the sum of leptonic momenta is 
$q$ = $q^{\phantom{l}}_1+q^{\phantom{l}}_2$ =
$q^{\phantom{l}}_e+q^{\phantom{l}}_{\overline{\nu}}$.

By decoupling the spinor trace between the matrix element and the 
partonic process, and boosting $p^{\phantom{l}}_B$ 
to an infinite momentum frame,
the decay amplitudes in Eq.~(\ref{M-TH}) can be approximated as
($M^{\alpha\mu}_1$ and $M^{\alpha\mu}_2$ refer to the 
contributions from the direct and crossed graphs, respectively),
\begin{equation}
M^{\alpha\mu}_i
\approx
\int dx\;
T^{\sigma}_B(x) \;
H^{\alpha\mu}_{i\sigma}(p^{\phantom{l}}_b=x p^{\phantom{l}}_B, \;
p^{\phantom{l}}_{\Psi}, \; q)\; ,
\label{M-TH-x}
\end{equation}
where the matrix element is defined by
\begin{equation}
T^{\sigma}_B(x)
= \int \frac{p^{+}_B\; dy^-}{2 \pi}\;
\exp({i\, x \, p^{+}_B\; y^-})\;
\langle \; 0 \; \vert \; \overline{u}(0) \gamma^{\sigma}
\gamma^{\phantom{l}}_5 \;
           b(y^-) \; \vert \; B^- \rangle \; ,
\label{TB-x}
\end{equation}
and the partonic process $H$ is computed extracting the axial-vector 
component of the lowest order amplitudes $b\bar{u}\rightarrow
g^{\star} W^{\star}$.

Eq. (\ref{TB-x}) is given in terms of light-cone coordinates:
the $\pm$ components of any four-vector $k^\mu$ are defined as
$k^{\pm} = (k^0 \pm
k^3)/\sqrt{2}$. We also write $p^{\mu}_B$ = $\left(p^+_B, \;
{m^2_B} /(2 p^+_B),\; 0_{\perp}\right)$ using the boost-invariant
light-cone wavefunction language.

Due to the heavy quark mass, the typical virtuality of
$b$ and $\bar{u}$ in the $B$ meson should be much smaller than the
large momentum scale in ${H}$, and therefore we can
approximate the parton momenta $p^{\phantom{l}}_b$ and 
$p^{\phantom{l}}_{\bar{u}}$
in ${H}^{\alpha\mu}_{i\sigma}$ as 
\begin{equation}
p^{\phantom{l}}_b  = x\; p^{\phantom{l}}_B\; , \qquad
p^{\phantom{l}}_{\bar{u}}  = (1-x)\; p^{\phantom{l}}_B\; , \qquad
\mbox{with}\
x = {m^{\phantom{l}}_b \over m^{\phantom{l}}_B} \; .
\end{equation}
This approximation corresponds to dropping power corrections
suppressed by the hard momentum scale of the partonic process. Fixing
$x$ in ${H}^{\alpha\mu}_{i\sigma}$ 
we decouple the $x$ integration in Eq.~(\ref{M-TH-x}) to find
\begin{equation}
\int dx \; T^{\sigma}_B(x)
=
\langle \; 0 \; \vert \; \overline{u}(0) \gamma^{\sigma}
\gamma^{\phantom{l}}_5 \;
           b(0) \; \vert \; B^- \rangle \;
\equiv
i f^{\phantom{l}}_B\; p^{\sigma}_B \, .
\label{fB}
\end{equation}
For numerical estimates we use 
$f^{\phantom{l}}_B$ = 0.177 GeV
\cite{bern1,bern}\footnote{$f^{\phantom{l}}_B$ = 0.177 GeV
corresponds to the usual annihilation of a $b \bar u$ pair in the color
singlet configuration. In our case the $b \bar u$ pair is in a color
octet configuration and therefore we expect that our
$f^{\phantom{l}}_B$ is actually
smaller than 0.177. Our results for the perturbative production of
$J/\Psi$ in semileptonic $B$ decays can thus be considered to be an
upper bound.}. Substituting Eq.~(\ref{fB}) into Eq.~(\ref{M-TH-x}) we
obtain a factorized expression for the decay amplitude of
$B\rightarrow g^{\star} W^{\star}$,
\begin{equation}
M^{\alpha \mu}_i
\approx
(i f^{\phantom{l}}_B\; p^{\sigma}_B) \; \left.
H^{\alpha\mu}_{i\sigma}\left(p^{\phantom{l}}_b=xp^{\phantom{l}}_B, \;
p^{\phantom{l}}_{\Psi}, \; q\right)\right|_{x=\frac{m^{\phantom{l}}_b}{m^{\phantom{l}}_B}} \, .
\label{M-fac}
\end{equation}
The amplitudes in Eq.~(\ref{M-TH}) can thus be written as
\begin{eqnarray}
M^{\alpha \mu}_1
&=&
f^{\phantom{l}}_B\,
{{g^{\phantom{l}}_S g^{\phantom{l}}_W } \over
 {2 \sqrt{2} }}
{1 \over {(p^{\phantom{l}}_{\Psi}-p^{\phantom{l}}_{\bar{u}})^2 -m^2_{\bar{u}}}}\,
V^{\phantom{l}}_{ub} \left[
-i \; \varepsilon^{\gamma \delta \alpha \mu}
p_{B\gamma} (p^{\phantom{l}}_{\Psi}-p^{\phantom{l}}_{\bar{u}})_{\delta} \right.
\nonumber \\
&+& \left.
 p^{\mu}_B (p^{\phantom{l}}_{\Psi}-p^{\phantom{l}}_{\bar{u}})^{\alpha}
+p^{\alpha}_B (p^{\phantom{l}}_{\Psi}-p^{\phantom{l}}_{\bar{u}})^{\mu}
-p^{\phantom{l}}_B\cdot (p^{\phantom{l}}_{\Psi}-p^{\phantom{l}}_{\bar{u}})\, g^{\alpha\mu} \right] \; ,
\end{eqnarray}
and (for the crossed diagram),
\begin{eqnarray}
M^{\alpha \mu}_2
&=&
f^{\phantom{l}}_B\,
{{g^{\phantom{l}}_S g^{\phantom{l}}_W} \over
  2 \sqrt{2}}
{1 \over {(q-p^{\phantom{l}}_{\bar{u}})^2-m^2_b}} \,
V^{\phantom{l}}_{ub} \left[
-i \; \varepsilon^{\gamma \delta \alpha \mu}
p^{\phantom{l}}_{B\gamma} (p^{\phantom{l}}_{\bar{u}}-q)_{\delta} \right.
\nonumber \\
&+& \left.
 p^{\mu}_B (-p^{\phantom{l}}_{\bar{u}}+q)^{\alpha}
+p^{\alpha}_B (-p^{\phantom{l}}_{\bar{u}}+q)^{\mu}
-p^{\phantom{l}}_B \cdot (-p^{\phantom{l}}_{\bar{u}}+q)\, g^{\alpha\mu}\right] \; .
\end{eqnarray}
We denote the strong and weak coupling constants by 
$g^{\phantom{l}}_S$ and $g^{\phantom{l}}_W$ respectively. 

The $W^{\star}$ coupling to the lepton current is 
evaluated in the usual way. After squaring this contributes a factor 
\begin{eqnarray}
L^{\alpha\beta}
= g^2_W \; \left( 1 \over {q^2-m^2_W} \right)^2 \;
\left[i \varepsilon^{\gamma\delta \alpha \beta}
        q^{\phantom{l}}_{1 \gamma} q^{\phantom{l}}_{2 \delta}
    + q^{\beta}_1 q^{\alpha}_2
    + q^{\alpha}_1 q^{\beta}_2
    - q^{\phantom{l}}_1 \cdot q^{\phantom{l}}_2\, g^{\alpha \beta}
\right]
\label{lept}
\end{eqnarray}
to the decay rate. 

Our remaining task is to parametrize the production of 
a $J/\Psi({}^3 S_1)$ state from the 
$c \overline{c}$ pair. We do this with projector techniques  
and compute the overall color factor below
(in Eq. (\ref{col}) \cite{mal}). We write 
\begin{eqnarray}
J_{\mu \nu}(p^2_{\Psi}) = \sum_{\lambda} \left(-i \over
p^2_{\Psi}\right) {\rm Tr}[(-i
g^{\phantom{l}}_S \gamma^{\phantom{l}}_{\mu}) \; \Pi^{\rho}_1] \;
\varepsilon^{(\lambda)}_{\rho}(p^{\phantom{l}}_{\Psi}) \cdot
\left(i \over p^2_{\Psi}\right) {\rm Tr}[(i
g^{\phantom{l}}_S \gamma^{\phantom{l}}_{\nu}) \; \Pi^{*\sigma}_1] \;
\varepsilon^{*(\lambda)}_{\sigma}(p^{\phantom{l}}_{\Psi}) \; ,
\end{eqnarray}
with the projector defined as \cite{mal}
\begin{eqnarray}
\Pi^{\lambda}_1 = {1 \over {\sqrt{m^3_{\Psi}}}} {1 \over 4}
\left(\not{p^{\phantom{l}}_{\Psi}} - m^{\phantom{l}}_{\Psi}\right)
\gamma^{\lambda} \left(\not{p^{\phantom{l}}_{\Psi}} +
m^{\phantom{l}}_{\Psi}\right) \; .
\end{eqnarray}
The total contribution to the decay rate from the 
$c \overline{c}$ line is then
\begin{eqnarray}
J_{\mu \nu}(p^2_{\Psi}) = 4\, g^2_S \left(1 \over
p^2_{\Psi}\right)^2 {1 \over {m^{\phantom{l}}_{\Psi}}} \left[p^{\Psi}_{\mu}
p^{\Psi}_{\nu} - m^2_{\Psi} g^{\phantom{l}}_{\mu \nu}\right] \; ,
\label{ccline}
\end{eqnarray}
and the overall color factor is
\begin{eqnarray}
{1 \over {N^2_c-1}} \sum_{ABC} {\rm Tr}[T^A C^B_8] \; {\rm Tr}[T^A C^C_8] = {1 \over
2}\; , \qquad C^A_8 = \sqrt{2}\; T^A\; ,
\label{col}
\end{eqnarray}
where $T^A$, A = 1...8, are the generators of the color group.

As shown in Eq.~(\ref{fac-NRQCD}), the rate of the process
$B^- \rightarrow J/\Psi\, e^- \overline{\nu}_e \, X$ includes the
NRQCD matrix element $\langle {\cal O}^{J/\Psi}_8 ({}^3 S_1) \rangle$
defined in \cite{mal,ben,braat}, which represents the probability for
an almost point-like $c\bar{c}$ pair in a color-octet $^3S_1$ state
to bind and form a $J/\Psi$. The matrix element is nonperturbative
and has to be extracted from experimental data, and
its numerical value is \cite{mal,ben}:
\begin{eqnarray}
\langle {\cal O}^{J/\Psi}_8 ({}^3 S_1) \rangle
= 1.19 \times 10^{-2} \; {\rm GeV}^3\; .
\label{matel}
\end{eqnarray}
As a final step to obtain the decay rate we need to divide by 
the number of color and polarization states of the $c\bar{c}({}^3 S_1)$ 
produced, $N_{col}N_{pol} = 24$, as explained in Ref.~\cite{mal}. 

Collecting our results, we obtain for the matrix element squared 
\begin{eqnarray}
\vert {\cal M} \vert^2 = \left(M^{\alpha \mu}_1+M^{\alpha
\mu}_2\right)^{\star}
\left(M^{\beta \nu}_1+M^{\beta \nu}_2\right)
L^{\phantom{l}}_{\alpha \beta}
\; J^{\phantom{l}}_{\mu \nu}(p^2_{\Psi}) \; {1 \over 2}\; {{\langle {\cal
O}^{J/\Psi}_8 ({}^3 S_1) \rangle} \over {N_{col}N_{pol}}} \; ,
\label{exclus}
\end{eqnarray}
which includes the color factor $1/2$ obtained in Eq. (\ref{col}). 
In our numerical results we use the 
boson masses \cite{pdg} 
$m^{\phantom{l}}_B$ = 5.279 GeV, $m^{\phantom{l}}_{\Psi}$ =
3.097 GeV, and the current quark mass $m^{\phantom{l}}_b$ =
4.9 GeV, together with $\vert V^{\phantom{l}}_{ub} \vert$ = 0.0035 \cite{bur},
$\alpha^{\phantom{l}}_S(2 m^{\phantom{l}}_c)$ = 0.266 \cite{gly} and 
$\tau^{\phantom{l}}_{B^-}$ = 1.653 $\times$ 10$^{-12}$ sec \cite{pdg}.
We find a branching ratio:
\begin{equation}
{\rm BR}(B^- \rightarrow J/\Psi \; e^- \overline{\nu}_e \, X) = 9 \times
10^{-11}\; .
\label{pert3}
\end{equation}

\begin{figure}[!htb]
\begin{center}
\epsfxsize=10cm
\centerline{\epsffile{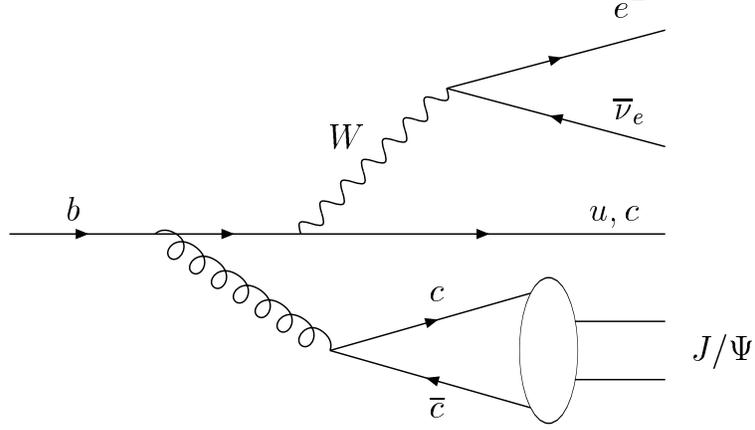}}
\end{center}
\caption{Spectator graph analogous to Fig. \ref{fig:graph1} for $B$
$\rightarrow$ $J/\Psi$ $e^-
\overline{\nu} \; X_{u,c}$. The gluon propagator may also be attached
to the outgoing quark $u$ or $c$.}
\label{fig:graph11}
\end{figure}

The process in Fig.~\ref{fig:graph1} which we have calculated
is the simplest one giving rise to 
$B^- \rightarrow J/\Psi \; e^- \overline{\nu}_e \; X$. However, it suffers 
from a suppression factor $f^{\phantom{l}}_B$ associated with the
$b\bar{u}$ annihilation.
For this reason it is possible that a process involving more quarks but 
no annihilation might actually be larger. To investigate this possibility 
we now estimate the rate of the process 
$B \rightarrow J/\Psi \; e^- \overline{\nu}_e \; X_{u,c}$
as in Fig.~\ref{fig:graph11}. 

We adopt once more the NRQCD factorization form of 
Eq. (\ref{fac-NRQCD}) to calculate separately 
the matrix elements of $b$ $\rightarrow$ $q$ $g^{\star} e^-
\overline{\nu}_e$
($q$ = $u,c$), and for the conversion of the $c\overline{c}$ pair 
into a $J/\Psi$.  We denote by $H^{\alpha\mu}_1$ and
$H^{\alpha\mu}_2$ the amplitudes of the partonic processes 
where the gluon is emitted from the $b$ and $q$ lines respectively.  
The squared partonic matrix element is then
\begin{eqnarray}
M^{\alpha\mu\beta\nu} = {\rm Tr}\left[(H^{\alpha\mu}_1 + H^{\alpha\mu}_2)
(H^{\beta\nu}_1+H^{\beta\nu}_2)^{\dagger}\right],
\end{eqnarray}
with 
\begin{eqnarray}
H^{\alpha\mu}_1 &=&
{{g^{\phantom{l}}_S g^{\phantom{l}}_W} \over {2 \sqrt{2} }}
{1 \over {(p^{\phantom{l}}_b-p^{\phantom{l}}_{\Psi})^2 -m^2_b}}\,
V^{\phantom{l}}_{uq} \overline{u}^{\phantom{l}}_q(p^{\phantom{l}}_q)
\gamma^{\alpha} (1-\gamma^{\phantom{l}}_5)
(\not{p}^{\phantom{l}}_b-\not{p}^{\phantom{l}}_{\Psi}) \gamma^{\mu}
u^{\phantom{l}}_b(p^{\phantom{l}}_b)\;, \nonumber \\
H^{\alpha\mu}_2 &=&
{{g^{\phantom{l}}_S g^{\phantom{l}}_W} \over {2 \sqrt{2} }}
{1 \over {(p^{\phantom{l}}_b-q)^2 -m^2_q}}\,
V^{\phantom{l}}_{uq} \overline{u}^{\phantom{l}}_q(p^{\phantom{l}}_q)
\gamma^{\mu} (\not{p}^{\phantom{l}}_b-\not{q})
\gamma^{\alpha} (1-\gamma^{\phantom{l}}_5)
u^{\phantom{l}}_b(p^{\phantom{l}}_b)\; .
\end{eqnarray}
The leptonic and hadronic $c\overline{c}$ parts and the NRQCD matrix
element are the same as for $B^- \rightarrow
J/\Psi\; e^- \overline{\nu}_e \, X$,
and have been given in Eqs. (\ref{lept}), (\ref{ccline}), and (\ref{matel}),
respectively. The overall color factor is different,
\begin{eqnarray}
{1 \over {N^{\phantom{l}}_c}} \sum_{ABC} {\rm Tr}[T^A C^C_8] \; {\rm
Tr}[T^B C^C_8]  \; \sum_{ij} \left(T^A\right)_{ij} \left(T^B\right)_{ij} = 
{2 \over3}\; , \qquad C^A_8 = \sqrt{2}\; T^A\; .
\label{col2}
\end{eqnarray}
All these factors are combined into the squared matrix element,
\begin{eqnarray}
\vert {\cal M} \vert^2 =
M^{\alpha\mu\beta\nu} L^{\phantom{l}}_{\alpha \beta}
\; J^{\phantom{l}}_{\mu \nu}(p^2_{\Psi}) \; {2 \over 3}\; 
{\langle {\cal O}^{J/\Psi}_8 ({}^3 S_1) \rangle \over N_{col}N_{pol}} \; .
\end{eqnarray}
Using the masses already defined plus \cite{pdg}
$m^{\phantom{l}}_c$ = 1.25 GeV and $\vert V^{\phantom{l}}_{bc} \vert$ = 0.041,
and a $b$ quark lifetime
$\tau^{\phantom{l}}_{b}$ = 1.6 $\times$ 10$^{-12}$ sec we find:
\begin{equation}
{\rm BR}(B \rightarrow J/\Psi \; e^- \overline{\nu}_e \; X_q) =
\cases{1 \times 10^{-12}\ \ {\rm for} \ q = c \; , \cr
4 \times 10^{-12}\ \ {\rm for} \ q = u \; .}
\label{pfour}
\end{equation}
These results show that the diagrams in Fig.~\ref{fig:graph1} generate 
the leading perturbative contribution to the process 
$B \rightarrow J/\Psi \; e^- \overline{\nu}_e \; X_q$, which is 
unobservably small.

\subsection{Total inclusive \mbox{\boldmath
$B$}-to-\mbox{\boldmath $c\bar{c}$} semileptonic decay}

The inclusive semileptonic rate for $B^- \rightarrow c \bar{c} \; e^- 
\overline{\nu}_e \, X$ is obtained from the diagrams in Fig.~\ref{fig:graph1} 
with some simple replacements. The calculation proceeds as in 
Section~\ref{secnrqcd} up until Eq.~(\ref{lept}). After that we simply 
couple the $g^{\star}$ to a $c\bar{c}$ line in the usual way, obtaining
\begin{eqnarray}
J^{\prime}_{\mu \nu}(p^{\phantom{l}}_{1},p^{\phantom{l}}_{2})
= 4\, g^2_S \left(1 \over
p^2\right)^2 \left[p^{\phantom{l}}_{1 \mu} p^{\phantom{l}}_{2 \nu} +
p^{\phantom{l}}_{2 \mu} p^{\phantom{l}}_{1 \nu} - (p^2
/2) g^{\phantom{l}}_{\mu \nu}\right] \; ,
\label{ccpline}
\end{eqnarray}
where $p^{\phantom{l}}_1$ and $p^{\phantom{l}}_2$
are the momenta of $c$ and $\bar c$, respectively, and $p$ =
$p^{\phantom{l}}_1$ + $p^{\phantom{l}}_2$.
The overall color factor now takes into account both the color octet
and the color singlet configurations of the incoming quarks $b$ and
$\bar u$, and can be calculated to be 2/9.
The squared matrix element becomes
\begin{eqnarray}
\vert {\cal M} \vert^2 = \left(M^{\alpha \mu}_1+M^{\alpha
\mu}_2\right)^{\star}
\left(M^{\beta \nu}_1+M^{\beta \nu}_2\right)
L^{\phantom{l}}_{\alpha \beta}
\; J^{\prime}_{\mu \nu}(p^{\phantom{l}}_{1},p^{\phantom{l}}_{2}) \; {2
\over 9}\; ,
\label{inclus}
\end{eqnarray}
where $M^{\alpha \mu}_i$ and $L^{\phantom{l}}_{\alpha \beta}$ are
the same quantities already used in Eq.~(\ref{exclus}).

\begin{figure}[!htb]
\begin{center}
\epsfxsize=12cm
\centerline{\epsffile{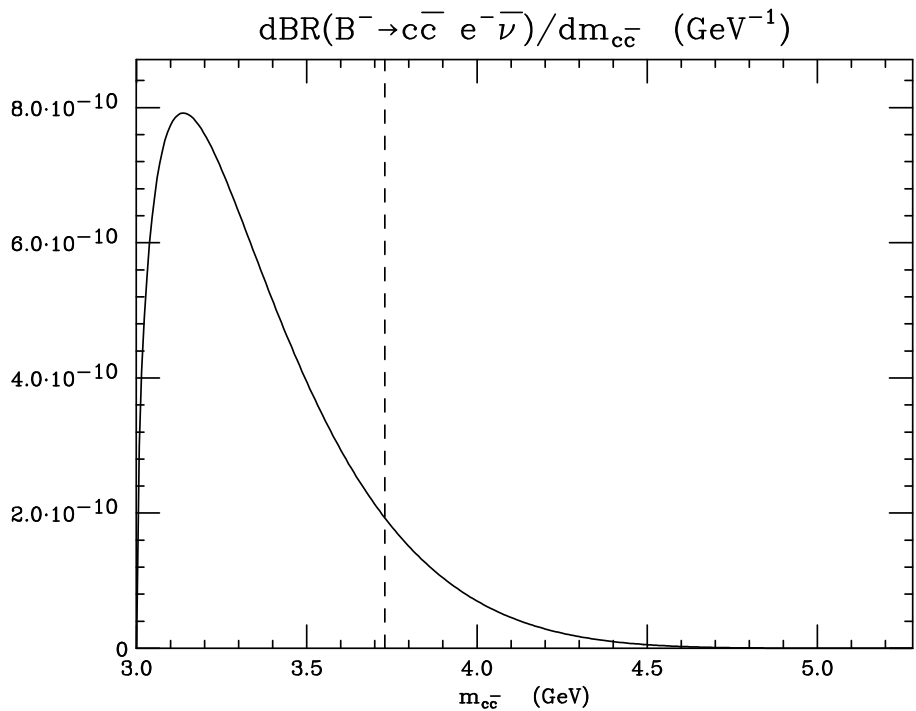}}
\end{center}
\caption{The differential branching ratio of $B^- \rightarrow c \bar
c \; e^- \overline{\nu} \, X$ is plotted vs. the invariant mass $m_{c \bar
c}$ of the $c {\bar c}$ pair. The vertical dashed line shows the energy
threshold for $D \bar D$ production.}
\label{fig:diff}
\end{figure}

We use the masses and CKM parameters defined previously, but we require 
the invariant mass of the $c\bar{c}$ pair to be larger than 3~GeV in
order to produce a physical $c \bar c$ meson. We obtain a branching ratio
\begin{eqnarray}
{\rm BR}(B^- \rightarrow c \bar{c} \; e^-
\overline{\nu}_e \, X) = 4.2 \times 10^{-10}. 
\end{eqnarray}
The differential branching ratio is plotted in Fig.~\ref{fig:diff} as a 
function of the $c\bar{c}$ pair invariant mass.
{}From Fig.~\ref{fig:diff} we can see that about 90\% of the time the 
$c\bar{c}$ pair will become a charmonium resonance and only about 10\% of 
the time it will lead to $D\bar{D}$ production.

\subsection{Color evaporation model}

According to the Color Evaporation model of $J/\Psi$ production, all
the $c\bar{c}$ pairs with invariant mass below open-charm threshold
should become charmonia by radiating soft gluons to neutralize the
color and other quantum numbers \cite{CEM}. The formation process
from the $c\bar{c}$ pairs to $J/\Psi$ mesons is assumed to be
independent of the color and invariant mass of the pairs. This is 
equivalent to choosing
\begin{equation}
F_{[c\bar{c}]_n \rightarrow J/\Psi}(q^2) 
= f^{\phantom{l}}_{J/\Psi}\, \theta(4m_D^2-4m_c^2-q^2)\, ,
\label{F-cem}
\end{equation}
in Eq.~(\ref{fact}). Here $4m_D^2$ represents the open-charm threshold, and 
$f^{\phantom{l}}_{J/\Psi}$ is a universal constant encoding the relative 
probability for a $c\bar{c}$ pair to become a $J/\Psi$ instead of 
other charmonium states, and needs to be fixed by data. 
Substituting Eq.~(\ref{F-cem}) into Eq.~(\ref{fact}), we derive the
formula for the decay rate in the Color Evaporation model,
\begin{equation} 
\Gamma^{\phantom{l}}_{B^- \rightarrow J/\psi \; e\bar{\nu} X}
= f^{\phantom{l}}_{J/\Psi}\ 
  \int_{2m_c}^{2m_D} dm_{c\bar{c}} \left[
  \frac{d\Gamma_{B^- \rightarrow c\bar{c} \;e\bar{\nu} X}
(m_{c\bar{c}})}{dm_{c\bar{c}}}
   \right]\; ,
\label{fac-cem}
\end{equation}
where $m_{c\bar{c}}$ is the invariant mass of the $c\bar{c}$ pair. 
Using the results of our previous section shown in Fig.~\ref{fig:diff}, 
this becomes
\begin{equation}
{\rm BR}(B^- \rightarrow J/\psi \; e\bar{\nu} \, X)
= f^{\phantom{l}}_{J/\Psi}\ 3.8 \times 10^{-10} \;.
\label{cembr}
\end{equation}
Comparing Eq.~(\ref{cembr}) with Eq.~(\ref{pert3}) we see that the CEM model 
result would agree with the NRQCD result if $f^{\phantom{l}}_{J/\Psi}
\sim 1/4.3$. This
fits very well with the expectation that $1/f^{\phantom{l}}_{J/\Psi}
\sim 2{~\rm~to~}7$
in Ref.~\cite{CEM}.

Our results in Eqs.~(\ref{pert3})~and~(\ref{pfour}) show that the 
perturbative QCD contributions to the processes 
$B^- \rightarrow J/\Psi \; e^- \overline{\nu}_e \, X$ and 
$B \rightarrow J/\Psi \; e^- \overline{\nu}_e \; X_{u,c}$ 
are too small to be observed by  near-future experiments. 
Qualitatively, these results are so small because they
involve two off-shell propagators as well as suppressions from
powers of $\alpha^{\phantom{l}}_S$ and from
the octet matrix element $\langle {\cal O}^{J/\Psi}_8 ({}^3 S_1) \rangle$. 
This leaves open the possibility of significant enhancements due to 
IC which we explore in the following section.

\section{The intrinsic charm contribution}\label{intic}

\begin{figure}[!htb]
\begin{center}
\epsfxsize=12cm
\centerline{\epsffile{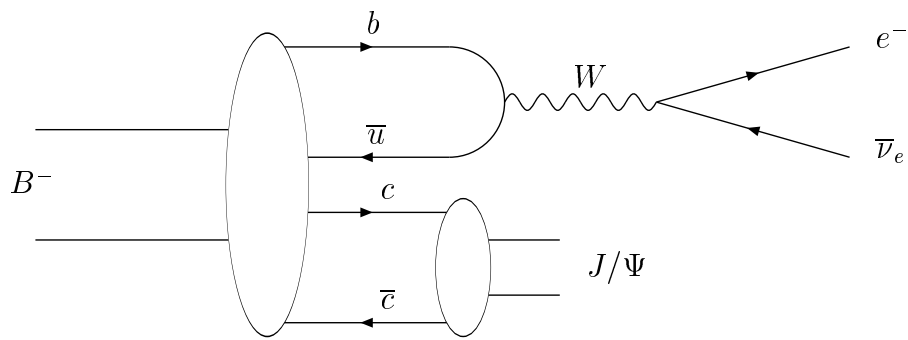}}
\end{center}
\caption{IC contribution to the decay $B^- \rightarrow J/\Psi \; e^-
\overline{\nu} \, X$.} 
\label{fig:graph2}
\end{figure}

The mechanism by which IC in a $B$ meson leads to the decay $B^-
\rightarrow$ $J/\Psi$ $e^- \overline{\nu}_e \, X$ is different from
the perturbative case. The two (anti)charm quarks are now part of
the initial wavefunction as shown in Fig. \ref{fig:graph2}.
In this picture a fluctuation in an ordinary $B^-$ meson produces
a nearly on-shell $c \overline{c}$ pair which is part of the initial 
wavefunction. There is a small probability for this fluctuation to 
occur \cite{IC}, but once it happens, it is possible for the 
off-shell $b\bar{u}$ quarks to annihilate leaving a $J/\Psi$ behind. 
In this section we attempt to estimate the rate of this process in the 
following way. We start from a $B^-$ meson wavefunction in the higher 
Fock state $|b\bar{u}c\bar{c}\, \rangle$ with a certain momentum distribution 
as proposed in the IC picture of Brodsky {\it et al.} \cite{IC}. We then
estimate the
annihilation rate of the $b\bar{u}$ components treating the $c\bar{c}$ 
as spectators. This annihilation is negligible unless the $b\bar{u}$ 
system is in a $J=1$ state, and in this case we think of it as an 
off-shell $B^{\star}$. As we will see, the invariant mass of the 
remaining $c\bar{c}$ pair is dominantly in the charmonium resonance 
region, so we assume that a certain fraction produces a $J/\Psi$ as in 
the CEM model discussed in the previous section.

The momentum fraction distribution of the $\vert b \bar u c \bar c \,
\rangle$ light-cone Fock component of $B^-$ implies that 
IC carries large momentum fractions,
unlike the small $x$ of extrinsic quarks from gluon
splitting. It is given by \cite{IC}:
\begin{equation}
{dP_{IC}^B \over dx_1 \cdots dx_4} = 
\ N_4 \, {\delta(1-\sum_{i=1}^4 x_i) 
                     \over
  (m_B^2 - \widehat{m}_{\bar u}^2/x_{\bar u}
         - \widehat{m}_b^2/x^{\phantom{l}}_b
         - \widehat{m}_c^2/x^{\phantom{l}}_c
         - \widehat{m}_{\bar c}^2/x^{\phantom{l}}_{\bar c})^2} \; .
\label{momf}
\end{equation}
In Eq. (\ref{momf}) the $x^{\phantom{l}}_i$ are the light-cone
momentum fractions
$(k^0+k^3)_i/(p^0_B+p^3_B)$, and $\widehat{m}_i$ = $(m^2_i+\langle
k^2_{\perp} \rangle_i)^{1/2}$ are the transverse masses of each
constituent as in Refs. \cite{IC}. In our
calculation we use the numerical values $\widehat{m}_b$ = 5 GeV,
$\widehat{m}_{u,d}$ = 0.45 GeV and $\widehat{m}_c$ = 1.8 GeV
\cite{IC,bhv}.
Integrating Eq. (\ref{momf}) over all possible momenum fractions
gives the probability $R_{IC}$ of having the $B$ meson in an IC
configuration. In terms of this $R_{IC}$ we obtain the four-particle 
normalization factor $N_4$ = $R_{IC}$ $\times$ 10$^5$ GeV$^4$. 
It has been suggested in Ref. \cite{bg} that IC effects in 
$B$ mesons may be four times larger than in the proton, where it is
estimated that they can occur at the 1\% level \cite{hsv}. In particular, 
this means that $R_{IC}$ could be as large as 0.04 according to 
Ref. \cite{bg}.

We are interested in a process in which the $c \overline{c}$ is in 
the particular configuration ${}^3 S_1$ color singlet state. 
There is, of course, a model dependent combinatorial factor 
associated with this. Because we lack a detailed dynamical model 
for IC, we will instead resort to a phase-space model 
similar in spirit to the CEM discussed in the previous section. 

We begin our estimate of the amplitude by treating the charm (anti)quarks as
spectators.  The wavefunction is then the product of a
($c\overline{c}$) state that generates the $J/\Psi$ and a virtual
($b\overline{u}$) state with the quantum numbers of  a $B^{*-}$
(other configurations are helicity suppressed) which decays into
leptons. This virtual ($b\overline{u}$) pair possesses an invariant mass
squared $m^2_{B^{\star}}$ = $q^2$ =
$(x^{\phantom{l}}_b+x^{\phantom{l}}_{\bar u})^2
m^2_B$. The amplitude squared reduces to the product of a 
leptonic part $L^{\phantom{l}}_{\mu\nu}$ which is the same 
as that in Eq. (\ref{lept}) and
the hadronic b--u line contribution. Combining them we find:
\begin{equation}
\vert {\cal M} \vert^2 = - f^2_{B^{\star}} m^2_{B^{\star}} {g^2_W
\over 8} \vert
V^{\phantom{l}}_{ub} \vert^2 g^{\mu \nu} L^{\phantom{l}}_{\mu\nu}\; = 
4 f^2_{B^{\star}} q^4 G^2_F \; \vert V^{\phantom{l}}_{ub} \vert^2\; .
\label{icmatel}
\end{equation}
It is evident from this result that
we treat the $b\overline{u}$ annihilation as the leptonic decay 
of an off-shell $B^\star$. For our numerical estimates we take  
$f^{\phantom{l}}_{B^\star}$ = $f^{\phantom{l}}_{B}$.

It remains to integrate Eq.~(\ref{icmatel}) over the phase space 
available to the lepton pair (determined by $m^2_{B^{\star}}$) and 
to convolute this result with the momentum distribution in the 
$B$ meson wavefunction, Eq. (\ref{momf}). In this convolution 
we impose kinematic constraints consistent with ``liberating'' the
IC pair which becomes a charmonium resonance or a $D \bar{D}$
pair.

\begin{figure}[!htb]
\begin{center}
\epsfxsize=12cm
\centerline{\epsffile{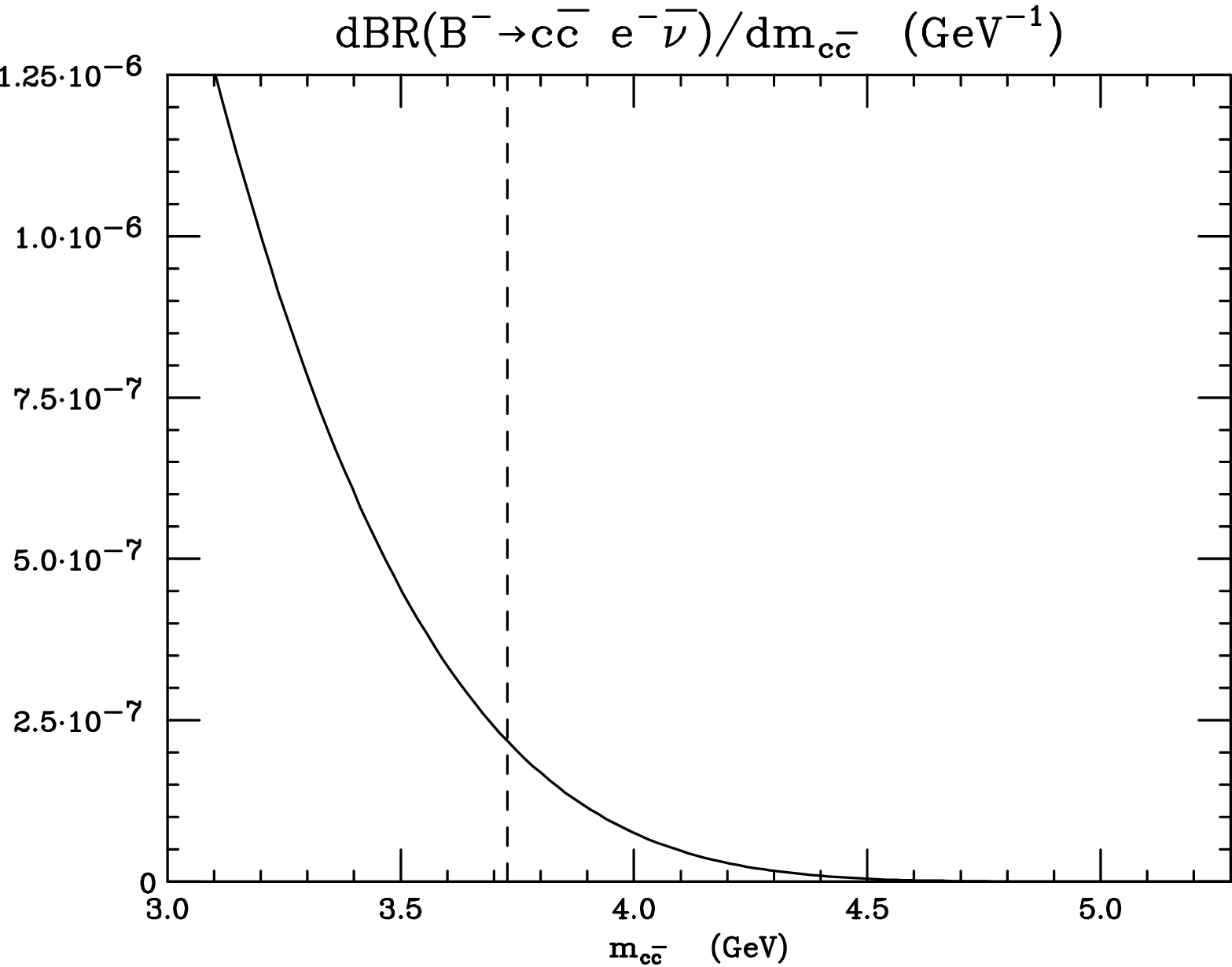}}
\end{center}
\caption{The IC contribution to the differential branching ratio of
$B^- \rightarrow c \bar c \; e^- \overline{\nu} \, X$ is plotted vs. the
invariant mass $m_{c \bar c}$ of the $c {\bar c}$ pair. The vertical
dashed line shows the energy threshold for $D \bar D$ production.}
\label{fig:diffic}
\end{figure}

Numerically, all this results in the differential decay rate 
shown in Fig.~\ref{fig:diffic}, where we have used $R_{IC}=0.04$ 
for illustration. 
The integrated rate leads to the inclusive branching ratio,
\begin{equation}
{\rm BR}^{\phantom{l}}_{IC}(B^- \rightarrow (c\bar{c}) \; e^-
\overline{\nu}_e\, X) = 
5 \times 10^{-7} \times {R_{IC} \over 0.04} \; .
\end{equation}
If instead we only integrate over the charmonium resonance region, we
obtain 
the estimate,
\begin{equation}
{\rm BR}^{\phantom{l}}_{IC}(B^- \rightarrow J/\Psi \; e^-
\overline{\nu}_e \, X) =
4 \times 10^{-7} \times {R_{IC} \over 0.04} \; .
\label{ic}
\end{equation}
Only a fraction of this result corresponds to $J/\Psi$, 
and if we use our perturbative 
calculations as guidance, this fraction could be of order 
$f_{J/\Psi} \sim 1/4.3$. However, as a signal for IC
any charmonium resonance that can be reconstructed (or for that matter any
$D \bar{D}$ channel) 
will serve. Integrating the distribution in Fig.~\ref{fig:diffic} above 
open-charm threshold leads to the estimate 
${\rm BR}^{\phantom{l}}_{IC}(B^- \rightarrow D \bar{D} 
\; e^- \overline{\nu}_e \, X) =
6 \times 10^{-8} \; R_{IC}/0.04 $.
If the fraction of IC in the $B$ meson can really be 
as large as $R_{IC}=0.04$, we see that semileptonic $B$ decay constitutes
a powerful probe of these ideas. Our result, Eq.~(\ref{ic}), is three
orders 
of magnitude larger than conventional, perturbative, mechanisms for 
producing charm pairs in semileptonic $B$ meson decay. This large 
enhancement can be understood qualitatively because the PQCD mechanism 
is suppressed with respect to the IC mechanism by two powers of 
$\alpha_S$, two propagators with an $m_\Psi$ mass scale, and the matrix 
element for $J/\Psi$ formation from a color octet $c\bar{c}$. At the same 
time the only relative suppression in the IC mechanism is the 
probability of finding the $B$ wave-function in the IC configuration, $R_{IC}$.
Roughly, the ratio of PQCD to IC induced rates goes as 
$\alpha_S^2 (\langle {\cal O}_8^{J/\Psi}(^3S_1)\rangle/m_\Psi^3) /R_{IC} 
\sim 10^{-3}$.

\begin{figure}[!htb]
\begin{center}
\epsfxsize=8.cm
\centerline{\epsffile{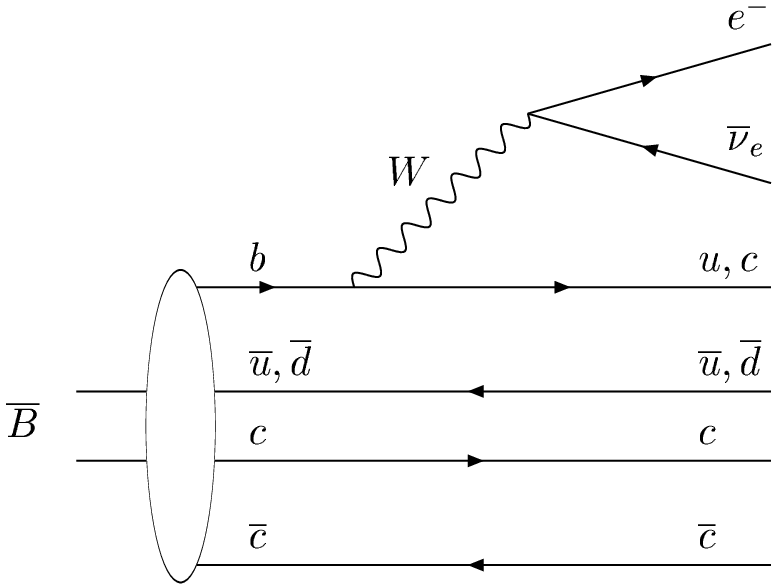}}
\end{center}
\caption{IC contribution to
the three-body spectator semileptonic decays $b \rightarrow J/\Psi$ $e^-
\overline{\nu}_e$ X$_{u,c}$.}
\label{fig:inc}
\end{figure}

We can perform a similar calculation for the decay $B \rightarrow
J/\Psi \; e^- \overline{\nu}_e \; X_{u,c}$, depicted in
Fig. \ref{fig:inc}. This process has at least one more particle in the 
final state and we might expect it to be suppressed by phase space 
with respect to $B \rightarrow J/\Psi \; e^- \overline{\nu}_e\,
X$. However, it does not suffer from the suppression factor associated
with the weak annihilation of the $b\bar{u}$ pair and for this reason we 
estimate it here. Our treatment of IC is similar to the previous channel. 
We calculate the semileptonic decay of an off-shell $b$-quark into 
three particles, $b \rightarrow e^- \overline{\nu}_e \; (u~{\rm or~}c)$  
and treat the remainder three quarks in the $B$ wavefunction 
as spectators. 

As in our previous example, we picture 
a nearly on-shell $c \overline{c}$ pair arising 
as a fluctuation in the $B$-meson wavefunction.
The semileptonic decay of the $b$-quark is standard, the only difference 
here being that the $b$-quark is off-shell and carries a fraction 
$x_b$ of the $B$ meson momentum. The partonic decay rate is thus 
(in terms of $m_b^\star = x^{\phantom{l}}_b m^{\phantom{l}}_B$):
\begin{equation}
\Gamma(b \rightarrow q \; e^- \overline{\nu}_e)
\approx {{G^2_F \vert V^{\phantom{l}}_{bq} \vert^2
{m^{\star}_b}^5} \over {192 \pi^3}} f\left(m^{\phantom{l}}_q \over
m^{\star}_b\right)\; , \quad {\rm q = u, c}\; ,
\end{equation}
where for the case of a $V^{\phantom{l}}_{bc}$ transition we also include the 
usual kinematic factor
\begin{equation}
f(x) = 1-8 x^2+8 x^6 -x^8 -24 x^4 \ln x.
\label{func}
\end{equation}
This factor is approximately equal to 1 for the $V^{\phantom{l}}_{bu}$
transition.
This result is now folded with the momentum distribution in the $B$ 
wavefunction, Eq.~(\ref{momf}), with kinematic constraints to once 
again ``liberate'' the IC pair. In spite of the CKM 
suppression $V_{ub}/V_{cb}$, the $V_{ub}$ mode can be as large or 
larger than the $V_{cb}$ mode because there is a very large
phase-space suppression producing three charm quarks. This suppression is 
so strong that the result is very sensitive to the value of $m_c$.
Using $m^{\phantom{l}}_c$ = 1.25 GeV
and a $b$ quark lifetime $\tau^{\phantom{l}}_b$ = 1.6 $\times
10^{-12}$ sec,  we estimate
\begin{equation}
{\rm BR}^{\phantom{l}}_{IC}(B \rightarrow J/\Psi \; e  \nu_e\;
X^{\phantom{l}}_q)
\approx \cases{1 \times 10^{-7} \times {R_{IC} / 0.04} \ \ {\rm for} \
q = c \; ,\cr 4 
\times 10^{-8} \times {R_{IC}/ 0.04} \ \ {\rm for} \ q = u \; .\cr}
\label{incl}
\end{equation}
If instead we use $m^{\phantom{l}}_c$ = 1.6 GeV we obtain
BR$^{\phantom{l}}_{IC}(b \rightarrow J/\Psi \; e^- \overline{\nu}_e\;
X^{\phantom{l}}_c)$ = 4 $\times$ 10$^{-9} \times {R_{IC}/ 0.04}$.
{}From this we conclude that the $V_{ub}$ mode is a small correction to our 
previous process so that $B^- \rightarrow J/\Psi \; e^- \bar{\nu} \; X_u$ 
is dominated by the annihilation mode  $B^- \rightarrow J/\Psi \; e^-
\bar{\nu} \, X$.
We also conclude that the channel with three charm quarks in the final state 
is smaller in spite of the CKM advantage it enjoys.

\section{Discussion and conclusions}\label{conc}

Up to now we have discussed the production of a $c\bar{c}$ state 
like $J/\Psi$ in a semileptonic $B^-$ decay. We have shown that 
the perturbative QCD contributions to this process are extremely 
small, giving rise to branching ratios at the $10^{-10}$ level.
Similar conclusions apply to the perturbative production of other
charmonium resonances or $D\bar{D}$ pairs.
We have also shown that there can be significantly larger contributions 
from IC components of the $B$-meson wavefunction.

It is simple to understand why the IC contribution to our
processes is much larger than the perturbative contribution. The
perturbative contribution suffers from several suppression factors
associated with the exchange of a perturbative gluon and the color
rearrangements that are needed to convert two color octet $q \bar q$
pairs into color singlet mesons. The IC contribution bypasses all
this because all necessary color exchanges and rearrangements take
place in a nonperturbative way inside the $B$-meson wavefunction. The
only suppression for this mechanism is then the small probability of
finding the $B$ meson in the IC configuration.

Adopting a simple spectator model and accepting recent suggestions in the 
literature as to the possible size of the IC component, we have estimated 
that $B^- \rightarrow (c\bar{c}) \; e^- \bar{\nu} X$ could have a branching 
ratio as large as $5 \times 10^{-7}$. We now show that at this level, 
the process $B^- \rightarrow (c\bar{c}) \; e^- \bar{\nu} X$ could be observed 
in the near future.

We start by estimating the sensitivity of CDF to the process 
$B \rightarrow J/\Psi \; e \nu X$ in Run II by comparing it to 
$B^{\phantom{l}}_c \rightarrow J/\Psi \; l \nu$, which they expect to see 
and which has the same final state. 
Assuming 2 fb$^{-1}$ and 15
fb$^{-1}$ respectively for the integrated luminosities during
Run IIa and Run IIa+Run IIb, and using \cite{teva}
\begin{equation}
{{\rm BR}(b \rightarrow B^{\phantom{l}}_c) \over {\rm BR}(b
\rightarrow B^-)} \times
{{{\rm BR}(B^{\phantom{l}}_c \rightarrow J/\Psi \; l \nu)} \over
{{\rm BR}(B^- \rightarrow J/\Psi \; K^-)}}
= 0.13 \pm 0.06 \;,
\end{equation}
leads to the estimate \footnote{We are grateful to
T.~LeCompte and M.~Tanaka for providing us with this estimate.}
\begin{equation}
N(B^{\phantom{l}}_c \rightarrow J/\Psi \; l \nu) \sim 800 \; {{\int
\cal{L}} \over {2 \; {\rm fb}^{-1}}} = 6000 \; {{\int
\cal{L}} \over {15 \; {\rm fb}^{-1}}} \;.
\end{equation}
{}From this estimate, with $l$ = $\mu$, $e$, the number of events for 
$B^- \rightarrow J/\Psi \; l \nu \, X$ is given by
\begin{equation}
N(B \rightarrow J/\Psi \; l \nu \, X) \sim 5 \times {{\int \cal{L}}
\over {15 \;
{\rm fb}^{-1}}} \times {{\rm BR}_{IC}(B \rightarrow J/\Psi \; l
\nu \, X) \over 10^{-7}} \; .
\label{ne}
\end{equation}
Therefore, CDF could be able to use $B \rightarrow J/\Psi \; l \nu \, X$ to 
test IC in the upcoming Run II. Notice that the open-charm channels 
such as  $B \rightarrow D^+ D^-\; l \nu \, X$ could also be used. Our modeling 
of IC predicts a smaller branching ratio, of order $6 \times 10^{-8}$. 
However, the identification of charged $D$ mesons could be up to 
10 times more efficient than that of $J/\Psi$ since the latter is
reconstructed from its leptonic decay which has a small branching ratio, 
$B(J/\Psi \rightarrow \mu^+\mu^- + e^+e^-) \approx 0.12$ \cite{pdg}. 

A similar projection for \babar\ can be based on the decay $B^0
\rightarrow D^{\star} \; l \nu$. Given the
7517 $\pm$ 104 reconstructed events for $B^0 \rightarrow D^{\star} \; l \nu$
with a total integrated luminosity of 9.7 fb$^{-1}$ \cite{BB}, and a projected
total luminosity of $\sim$ 542 fb$^{-1}$ for year 2006, we have:
\begin{equation}
N(B^0 \rightarrow D^{\star} \; l \nu) \sim 4.2 \times 10^5 \; {{\int
\cal{L}} \over {542 \; {\rm fb}^{-1}}} \; ,
\end{equation}
and from this we can estimate
\begin{eqnarray}
N(B \rightarrow J/\Psi \; e \nu\, X) &\sim& {\rm BR}(J/\Psi \rightarrow \;
e e + \mu \mu) \times N(B^0 \rightarrow D^{\star} \; l \nu) \times {{
{{\rm BR}_{IC}(B \rightarrow J/\Psi \; e \nu \, X)} \over {\rm BR}(B^0
\rightarrow D^{\star} \; l \nu)} } \nonumber \\
&\sim& 1 \times \; {{\rm BR}_{IC}(B \rightarrow J/\Psi \; e
\nu \, X) \over 10^{-6}} \; .
\end{eqnarray}
We have included a factor BR$(J/\Psi \rightarrow \;
e e + \mu \mu)$ to roughly account for the lower reconstructing
efficiency of $J/\Psi$ compared to $D^{\star}$. From these numbers
we conclude that ${\rm BR}_{IC}(B \rightarrow J/\Psi \; e
\nu \, X)$ could be probed at the 10$^{-7}$ level at CDF and at the
10$^{-6}$ level at \babar. Similar considerations can be applied to 
Belle. 

Near future experiments could use semileptonic $B$-meson decays to
probe the ideas of IC at an interesting level. Their study would
complement the suggestion of Ref.~\cite{hou} and can place limits on
the IC content of $B$ mesons.

\section*{Acknowledgments}

\noindent This work was supported in part by the U.S. Department of
Energy under Contract Numbers DE-FG02-01ER41155 and DE-FG02-87ER40371.
We also thank James Cochran, Thomas LeCompte, Masa Tanaka, Fabio Maltoni
and Aida El-Khadra for discussions.


\begin{thebibliography}{99}

\bibitem{IC}
S.~J.~Brodsky, P.~Hoyer, C.~Peterson and N.~Sakai,
Phys.\ Lett.\ B {\bf 93}, 451 (1980);
S.~J.~Brodsky, C.~Peterson and N.~Sakai,
Phys.\ Rev.\ D {\bf 23}, 2745 (1981).

\bibitem{hou}
C.-H.~V.~Chang and W.-S.~Hou, Phys. Rev. D {\bf 64},
071501 (2001).

\bibitem{alex}
A.~A.~Petrov, Phys. Rev. D {\bf 58}, 054004 (1998).

\bibitem{bg}
S.~J.~Brodsky and S.~Gardner,
Phys.\ Rev.\ D {\bf 65}, 054016 (2002).

\bibitem{bk}
S.~J.~Brodsky and M.~Karliner, Phys. Rev. Lett. {\bf 78},
4682 (1997).

\bibitem{hsv}
B.~W.~Harris, J.~Smith, and R.~Vogt,
Nucl.\ Phys.\ {\bf B461}, 181 (1996).

\bibitem{BBL-NRQCD}
G.~T.~Bodwin, E.~Braaten, and G.~P.~Lepage,
Phys.\ Rev.\ D {\bf 51}, 1125 (1995);
Erratum {\it ibid.} \ {\bf 55}, 5853 (1997).

\bibitem{CEM}
H. Fritzsch, 
Phys.\ Lett.\ {\bf B67}, 217 (1977);
M. Gl\"{u}ck, J.F. Owens, and E. Reya,
Phys.\ Rev.\ {\bf D17}, 2324 (1978).


\bibitem{mal}
M.~Beneke, F.~Maltoni, and I.~Z.~Rothstein, Phys. Rev. D {\bf 59},
054003 (1999); F.~Maltoni, M.~L.~Mangano, and
A.~Petrelli, Nucl. Phys. {\bf B519}, 361 (1998).

\bibitem{bern1}
C.~W.~Bernard,
Nucl. Phys. (Proc. Suppl.) {\bf 94}, 159 (2001).

\bibitem{bern}
C.~Bernard {\it et al.},
Phys.\ Rev.\ D {\bf 65}, 014510 (2002).


\bibitem{ben}
M.~Beneke, Lectures given at the {\it 24th Annual SLAC
Summer Institute on Particle Physics: The Strong Interaction, From
Hadrons to Protons} (SSI 96), Stanford, CA, in ``Stanford 1996, The
strong interaction, from hadrons to partons'', 549-574.

\bibitem{braat}
E.~Braaten, S.~Fleming, and T.~C.~Yuan,
Annu. Rev. Nucl. Part. Sci. {\bf 46}, 197 (1996).

\bibitem{pdg}
Particle Data Group, D.~E.~Groom {\it et al.},
Eur.\ Phys.\ J.\ C {\bf 15}, 1 (2000).

\bibitem{bur}
A.~J.~Buras,
Introductory lecture given at {\it KAON2001: International Conference on CP
Violation}, Pisa, Italy, 12-17 Jun 2001, hep-ph/0109197.

\bibitem{gly}
G.~Eilam, M.~Ladisa, and Y.-D.~Yang,
Phys.\ Rev.\ D {\bf 65}, 037504 (2002).

\bibitem{bhv}
R.~Vogt, S.~J.~Brodsky, and P.~Hoyer,
Nucl.\ Phys.\ {\bf B360}, 67 (1991); {\it ibid.} {\bf 383}, 643 (1992). 

\bibitem{teva}
CDF Collab., K.~Anikeev {\it et al}., preprint FERMILAB-Pub-01/197, December 2001, hep-ph/0201071.

\bibitem{BB}
\babar\ Collab., B.~Aubert {\it et al.}, SLAC-PUB-8530,
BABAR-CONF-00-08, Aug 2000, Talk given at the {\it 30th International
Conference on High-Energy Physics}, Osaka, Japan, 27 Jul - 2 Aug 2000,
hep-ex/0008052.

\end{thebibliography}
\end{document}